## ACS OMEGA

Article

# An Impedance Sensing Platform for Monitoring Heterogeneous Connectivity and Diagnostics in Lab-on-a-Chip Systems

Chunjie Zhang, Yang Su, Siyi Hu, Kai Jin, Yuhan Jie, Wenshi Li, Arokia Nathan, and Hanbin Ma*

Cite This: *ACS Omega* 2020, 5, 5098−5104

Read Online

ACCESS |  Metrics & More | 📖 Article Recommendations | 🔘 Supporting Information

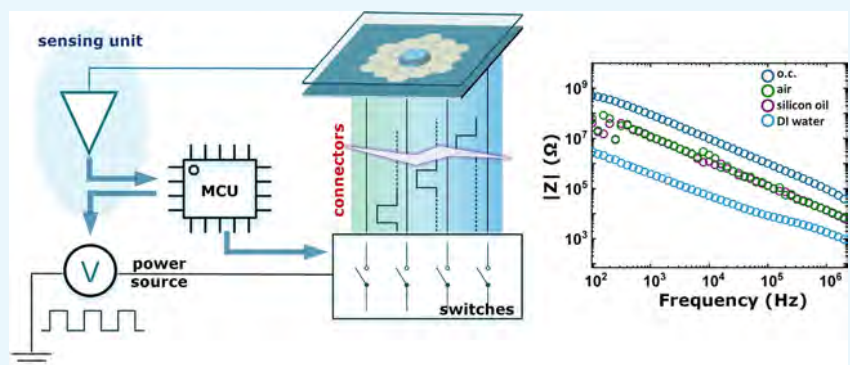

**ABSTRACT:** Reliable hardware connectivity is vital in heterogeneous integrated systems. For example, in digital microfluidics lab-on-a-chip systems, there are hundreds of physical connections required between a microelectromechanical fabricated device and the driving system that can be remotely located on a printed circuit board. Unfortunately, the connection reliability cannot be checked or monitored by vision-based detection methods that are commonly used in the semiconductor industry. Therefore, a sensing platform that can be seamlessly integrated into existing digital microfluidics systems and provide real-time monitoring of multiconnectivity is highly desired. Here, we report an impedance sensing platform that can provide fast detection of a single physical connection in timescales of milliseconds. Once connectivity is established, the same setup can be used to determine the droplet location. The sensing system can be scaled up to support multiple channels or applied to other heterogeneously integrated systems that require real-time monitoring and diagnostics of multiconnectivity systems.

## ■ INTRODUCTION

The use of single semiconductor technology routes, such as complementary metal-oxide semiconductors[1] (CMOSs) for high-speed circuits, insulated-gate bipolar transistors[2] (IGBTs) for high-power applications, and thin-film transistors (TFTs) for large-area and flexible electronics,[3] while they are necessary, can no longer be deemed sufficient in terms of requirements for complex systems in real-life applications. Heterogeneous electronics that integrate more than a single technology route is becoming the realistic engineering solution for realization of complex systems. Interconnection methods that reliably link different technology platforms are becoming key to successfully integrate heterogeneous systems. Here, reliability in connectivity is essential so as to achieve a long system lifetime.

Digital microfluidics (DMF) is a typical demonstration of a heterogeneously integrated system that consists of a microelectromechanical system (MEMS)-based microfluidic device, co-integrated with a CMOS-based peripheral circuits or drive electronics.[4−6] By feeding sequenced voltage signals through the peripheral circuits to an on-chip electrode array, a DMF chip can enable complex manipulation of each discrete droplet. The platform can support a wide range of applications,[7−9] which allows the possibility to enable true lab-on-a-chip (LOC) functions. One advantage of the DMF over other LOC methods is that no prefabricated microchannels are required, and the liquid droplets are manipulated over a two-dimensional region by electrostatic forces.[10−13] However, this requires that the droplets be controlled on top of the area where electrodes are physically located. A large number of electrodes, or an electrode array, is necessary for a complex biomedical procedure that requires simultaneous control of multiple liquid samples.[14] In most reported systems, the electrodes are individually connected to the driving electronics, and the number of physical connections increases as the number of electrodes scales up. Although the connection number can be reduced by introducing a thin-film electronics-based active matrix technique,[14,15] the increased device fabrication cost makes it difficult to be widely applied. In addition, a DMF device is often designed as a disposable part that needs to be replaced after a biochemical reaction due to biofouling.[16] Therefore, the reliability of the multiple connections between a DMF device and the driving electronics must be checked after every connection.

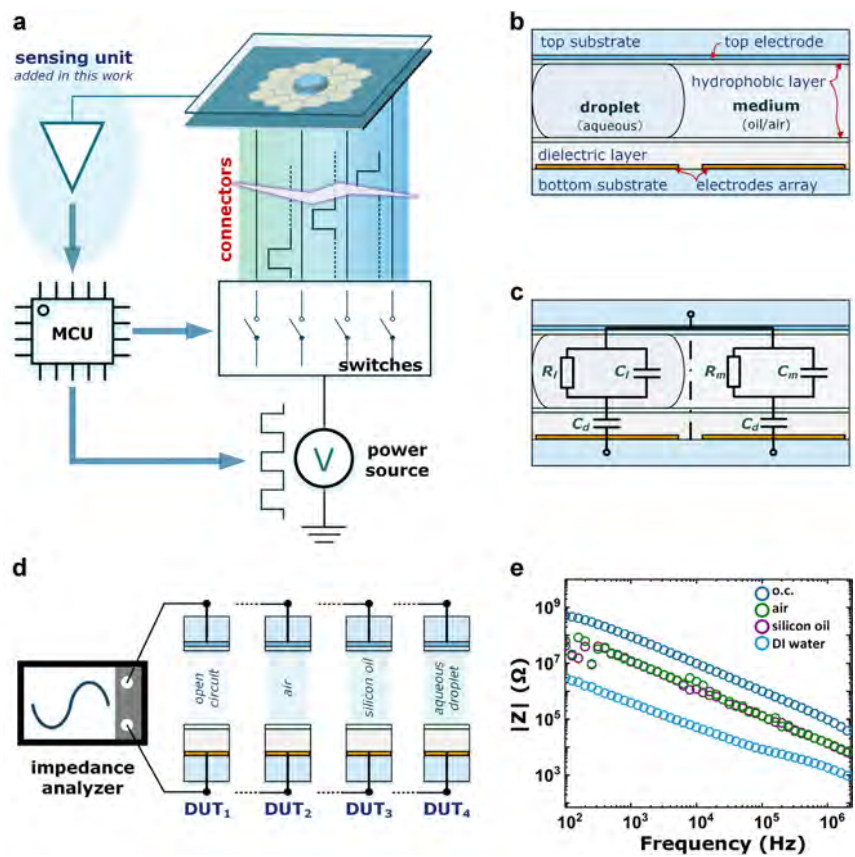

**Figure 1.** Digital microfluidic platform with the impedance-based sensing unit. (a) System diagram for conventional DMF; sensing unit is added between the top electrode from the device and the MCU. (b) Cross-sectional view of a double-plate DMF device. (c) Equivalent impedance models for the DMF device with the sample droplet and medium. (d) Impedance measurement setup for different DUTs. (e) Impedance measurement results.

Here, we report an impedance sensing technique for heterogeneously integrated systems, in which we demonstrate seamless integration of square-wave impedance sensing with a standard DMF system. For the first time, we realize real-time monitoring of the reliability of multiple connections between a microfluidics device and the driving electronics. In addition, by tuning the sensing parameters, we can use the same setup to detect the locations of the sample droplets. The system we report in this work has 180 connections using a flexible printed circuit board (FPC), and the system can be further upgraded to support more connections.

## ■ THEORY

**Impedance-Based Sensing Principles.** Some previous works also engaged impedance-based sensing on DMF devices.[17−19] As shown in Figure 1a, a typical DMF system consists of a microfluidics chip and the drive electronics. The two parts are physically connected by a multicontact connector. Different DMF device architectures are available, but the double-plate structure is the most commonly used for biochemical reactions.[20−23] The top plate has a common electrode, and the bottom plate hosts the electrode array. The driving electronics contain at least a power source, switching unit, and microcontroller (MCU). The MCU instructs the switches to feed power in a certain sequence, and the output of the switches is connected to each individual electrode on the bottom plate. The top plate in the conventional setup is grounded to provide zero potential to both the aqueous droplet and the surrounding medium. Figure 1b depicts the cross-sectional view of the device. The aqueous droplet sits on one electrode and is surrounded by the medium. Figure 1c shows the equivalent circuit for a device where the capacitance of the thin hydrophobic layer is neglected.

The detailed design and fabrication procedure of the DMF device can be found in the Experimental Methods section. We considered one bottom electrode and the top electrode as a two-terminal device under test (DUT). We performed four sets of impedance spectroscopy measurements, as shown in Figure 1d. The analyzer testing probes were unconnected to simulate the impedance of an open electrode connection (DUT1). When connected, the probes measured DUT2 (air medium), DUT3 (silicon oil medium), and DUT4 (aqueous droplet). Figure 1e depicts the impedance measurements for the four DUTs. All data show a pure capacitive behavior whereby the magnitudes of the measured impedance decrease linearly with increasing stimulus frequency. The aqueous droplet shows the smallest impedance value of all four DUTs due to the relatively large dielectric constant of the aqueous solution,[24] which is normally around 80. The impedance measurements for air and silicon oil are almost the same since the relative dielectric constant of silicon oil[25] is approximately 2, which is very close to that of air. For DUT1, which represents an open circuit, and the measured impedance indicates impedance analyzer's detection limit or noise floor. The results in Figure 1e clearly show that there are sufficient margins for the impedance sensing system to identify the status of the channel, that is, open circuit, connected with the medium and connected with aqueous droplets.

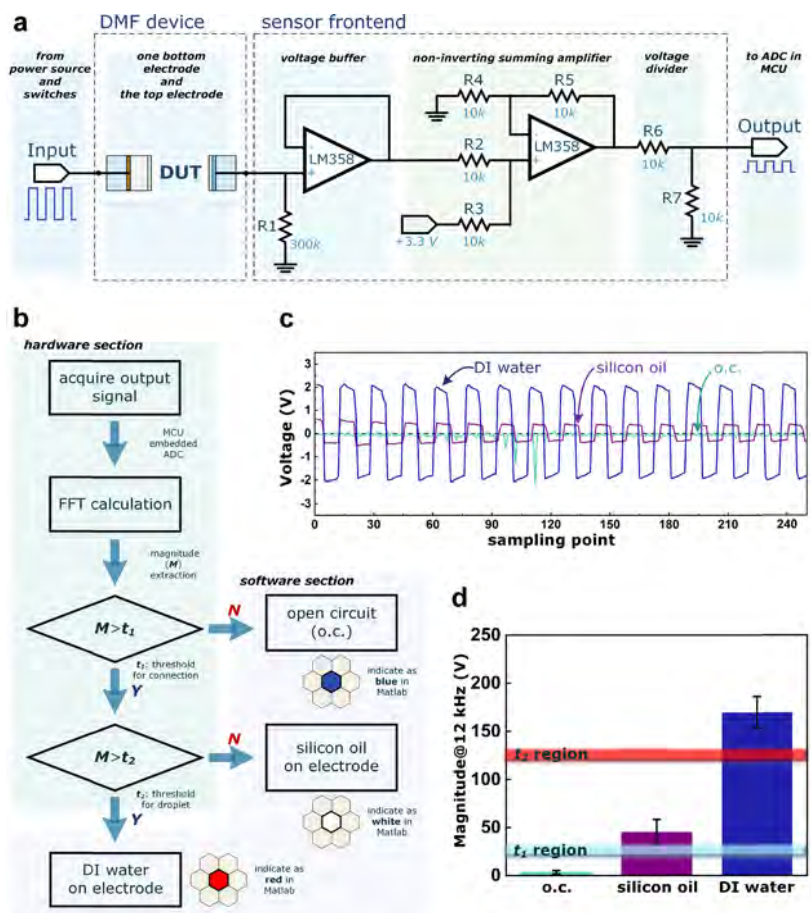

**Figure 2.** Sensing architecture. (a) Schematic diagram for the sensing unit. (b) Flow chart for the sensing steps, including the algorithm in MATLAB-based software. (c) ADC readout for DI water, silicon oil, and open circuit (o.c.). (d) Bar chart of the extracted magnitude values for DI water, silicon oil, and o.c.; possible regions for $t_1$ and $t_2$ are also indicated.

**Hardware and Software Design Considerations for Sensing.** We designed the driving electronics based on an STM32 development board with a customized power source and switches; please refer to the Experimental Methods section for the details. The driving system is able to control 180 channels with a DC voltage or a square wave up to 200 kHz. The voltage output range is 60–300 V. We used 280 V DC voltage to manipulate the sample, which turns out to be the best driving voltage on this platform. By using the same power source for DMF droplet manipulation, we generated a square wave (0 V to high V) as a stimulus in the impedance measurement for sensing. We connected the top electrode from the DMF device as the output terminal from the DUT. As shown in Figure 2a, we designed a sensor front-end circuit that consists of a voltage buffer, a non-inverting summing amplifier, and a voltage divider. The voltage buffer provides the sensed signal with enough driving ability for the stages that follow. Together with the equivalent capacitance from the

DUTs, an RC circuit is formed, which acts as a simple high-pass filter. The RC differentiator converts the input square wave step signal into an output containing both positive and negative elements. The non-inverting summing amplifier then levels up the output for the next stage. The voltage divider is connected to an analogue-to-digital converter (ADC), which is embedded in the MCU within the DMF driving electronics. By tuning the design parameters, the front end can convert a wide range of sensing signals to readable, meaningful data for the ADC.

Figure 2b shows the operational flow of the sensing unit, including both the hardware and software sections. After the output signal from the front end is acquired by the MCU-embedded ADC, we performed a fast Fourier transform (FFT) to extract the magnitude of the output signal. Figure 2c shows the sensing output waveforms for DUT1 (open circuit), DUT2 (connected with silicon oil as medium), and DUT3 (connected with a DI water droplet) under a 180 V 12 kHz stimulus (see Supplementary Figure 1 for stimulus signal optimization). As expected, the magnitude of DUT1 is much smaller than those of DUT2 and DUT3, which indicates high impedance. The measured DUT2 and DUT3 values also show a clear difference. This can be used to distinguish the droplet from the medium. We performed measurements for all connected electrodes. The results (see Supplementary Figure 2) show a high level of uniformity. We used STM32H7 in this work with embedded ADC at the 200 kHz sampling rate. We collected 256 sampling points to perform the FFT calculation, and the overall timing for one-channel detection was less than 2 ms. We used the value at 12 kHz (stimulus frequency) as the magnitude (M). If M is greater than a pre-set threshold value ($t_1$), the undertested channel has a secure connection between the electronics and the DMF device; otherwise, the electrode will be shaded blue in the software to indicate a bad connection. A second pre-set threshold value ($t_2$) is used to determine whether an aqueous droplet (red shaded in the software) or the medium is on top of the connected electrode. Figure 2d shows a bar chart for the average magnitude values of all 169 connections with three DUTs, clearly marking the possible margins for $t_1$ and $t_2$.

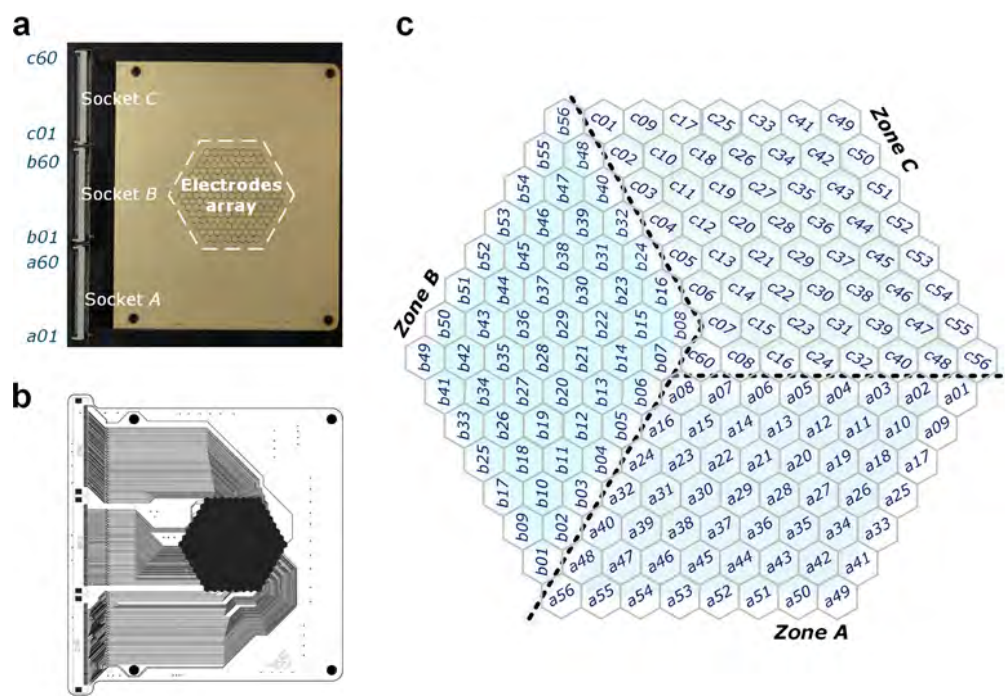

**Figure 3.** Electrode array. (a) Picture of the PCB-based electrode array with three 60-pin sockets for FPC connectors. (b) Layout of the PCB electrodes. (c) Electrode assignments into three zones; each zone is connected to a socket.

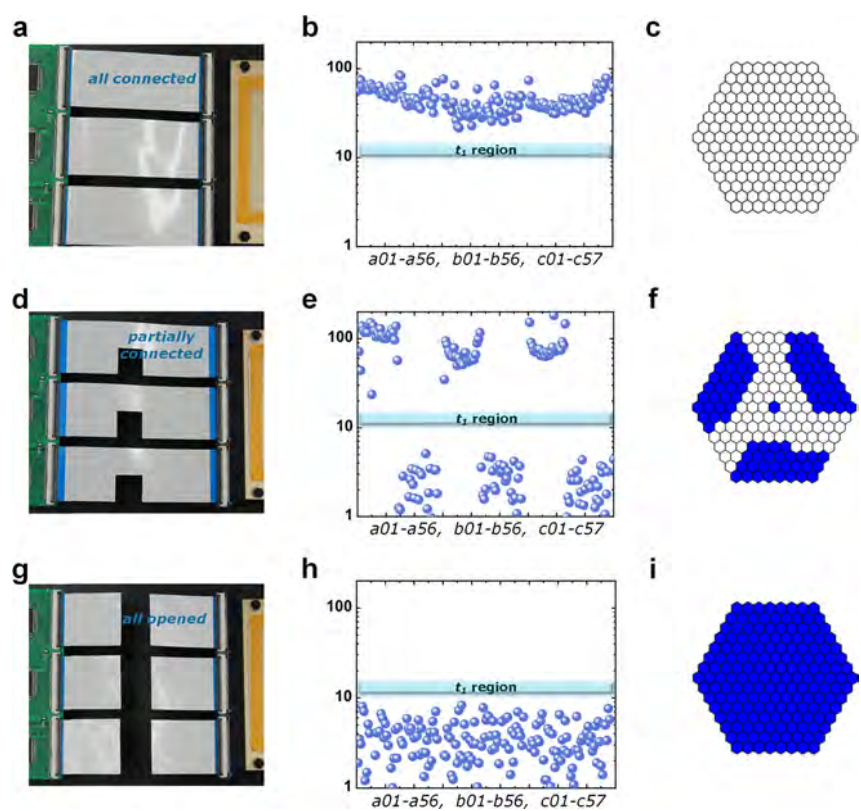

**Figure 4.** Connectivity measurements. (a) Picture of a fully connected system. (b) Measured magnitudes for 169 electrodes. (c) Indication by MATLAB for all connected electrodes. (d) Picture of a partially connected system, where half of the FPCs are intentionally cut off. (e) Measured magnitudes where half of them have a lower value. (f) Indication by MATLAB of bad connections (open circuit) by blue shading of the electrodes. (g) All connectors intentionally cut off. (h) Measured magnitudes show lower values for all electrodes. (i) All electrodes are shaded blue to indicate bad connections.

There are several advantages for using FFT to extract the impedance parameter in this work. First, we used the MCU to perform the calculations, eliminating the requirement of additional circuits or elements. The FFT calculation also shows a speed advantage where the single measurement time (including data acquisition) is less than 2 ms in our setup. In addition, FFT transforms signals from the time domain to the frequency domain. We select only the magnitude of the stimulus frequency. This method is able to filter all other elements that are mainly noise, therefore enhancing the sensing accuracy.

In this work, we designed an 8 × 8 × 8 hexagonal electrode array containing 169 electrodes (as shown in Figure 3a,b) with electrodes divided into three zones (A, B, and C) and connected to three 60-channel multiplexers. We also assigned addresses to each individual electrode (Figure 3c) to synchronize the hardware and software. A software interface was developed using a MATLAB platform, and a 169-electrode array was plotted. We used a color code to indicate the status of the different electrodes, in which blue indicates a bad connection (open circuit), white as idle showing a good connection with the medium on top of the channel, and red shows a good connection with an aqueous droplet.

## RESULTS AND DISCUSSION

**Connectivity Detection.** We used three FPCs to connect the electrode array with the driving electronics. Each FPC contains 60 individual channels. First, we connected all three connectors (Figure 4a) and performed a connection check. The 169 magnitudes are plotted in Figure 4b, where all the values are higher than $t_1$, indicating good connections. Then, the FPCs were cut (as shown in Figure 4d) to create bad connections (open circuit) for some of the channels. In Figure 4e, the corresponding measured magnitudes for the bad connections are below $t_1$. Figure 4f shows the indication of the bad connection electrodes on the MATLAB interface. Finally, we disconnected all three FPCs to leave all connections open (Figure 4g). All the measured data are below $t_1$, and all electrodes on the MATLAB were marked in blue for bad connections.

**Droplet Detection.** We used the same sensing unit to detect the droplet location in real time. Figure 5a–f shows single droplet movement and its real-time sensing results on the MATLAB. In our system, single channel detection takes less than 2 ms, and the overall detection time for 169 electrodes is approximately 300 ms. By adding a detection interval between each movement step, the exact droplet location can be determined. We performed the overall detection between each step, and all available electrodes were measured. This arrangement also supports multiple droplet detection. Figure 5g–m shows three droplet movements and the corresponding detection results. The videos for the experiments shown in Figure 5 are presented in the Supporting Information.

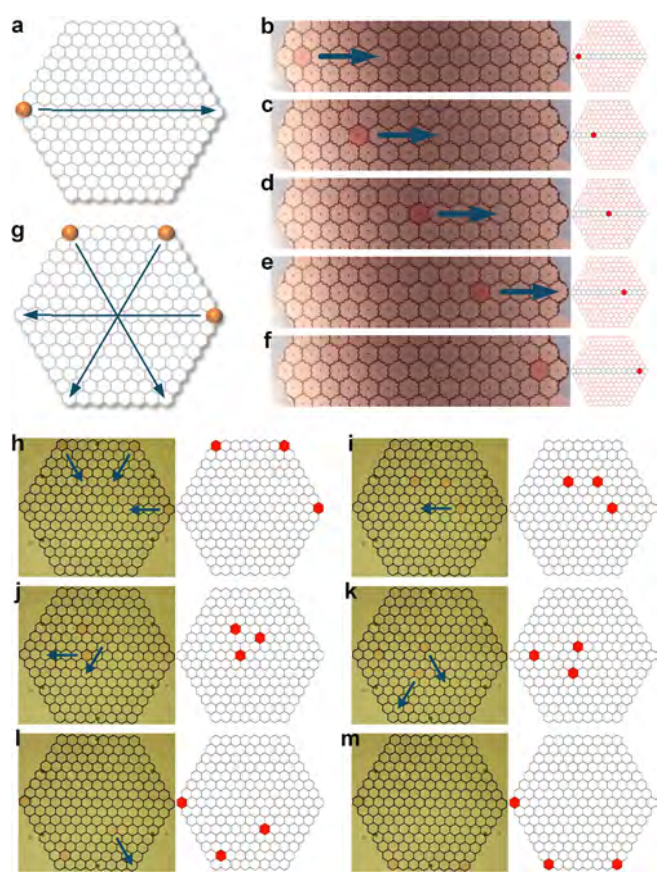

**Figure 5.** Real-time droplet location sensing. (a) Single droplet movement path design. (b–f) Single droplet movement with real-time feedback from the MATLAB interface. (g) Three droplet movement paths. (h–m) Three droplets moved with real-time feedback.

The FFT-based approach is engaged here to enable the potential to detect various kinds of samples or solution on the same chip. We tested seven different samples on our chip using different frequencies to find out the best frequency for detection as shown in Supplementary Figure 3. It turns out that DI water, ethanol, phosphate-buffered saline (PBS) solution (0.01 M), bovine serum albumin (BSA) solution (2 mg/mL), and NaCl solution (0.1538 M) show the highest sensitivity at 12 kHz, while oleylamine shows the highest sensitivity at 10 kHz, and octadecene shows the highest sensitivity at 30 kHz. Frequency multiplexing can be used to detect many droplets where various samples exist in a single chip. The resolution of frequency is 200 K/256 = 781.25 Hz, and the resolution can be increased by increasing sampling points.

## CONCLUSIONS

We have reported impedance sensing as a means for monitoring connectivity in an LOC system that can be fully integrated into existing DMF driving electronics. By measuring the impedance on the connectors, the system can not only recognize bad connections but also determine the locations of droplets. This work opens up a new frontier for dynamically monitoring the reliability of heterogeneously integrated systems of multiconnectivity with minimal effort. In addition, the impedance-based sensing platform shows the potential to be used in other heterogeneous integrated systems.

## EXPERIMENTAL METHODS

**DMF Device Fabrication.** We designed and fabricated double-plate DMF devices in this work. The bottom plate was a standard printed circuit board (PCB) with a hexagonal array of electrodes. Compared to conventional square patterns, a hexagonal pattern is closer to round shape, which would contribute less distortion for a liquid droplet while manipulated by the EWOD device. We designed 169 hexagonal electrodes and divided them into three zones. Three electrode zones were connected to three 60-pin FPC connectors. The electrode diameter was approximately 2 mm. On top of the electrodes, we coated a 50 $\mu m$-thick polyethylene (PE) film as the dielectric layer. Then, 0.9% Cytop (CTL-809M, AGC Chemicals) was used as a hydrophobic coating for both the bottom plate and an ITO-coated glass for the top plate. We defined a 200 $\mu m$ gap between the two plates using a Kapton spacer. We used 5 cSt silicon oil from Dow Corning as the oil medium and DI water with food dye for the aqueous droplets.

**Electronics System.** We used an STM32 by the ST NUCLEO-H743ZI Development Board as the MCU to control the customized power source and switches. We used three high-voltage solid-state multiplexers (HV507 from Microchips), each of which is able to control 60 channels. We used DMF software developed by ACXEL Tech Ltd., UK, for liquid manipulation, which can preprogram the moving path of the droplet. A MATLAB-based software interface was used for processing the sensing results.

## ASSOCIATED CONTENT

**ⓢ Supporting Information**

The Supporting Information is available free of charge at https://pubs.acs.org/doi/10.1021/acsomega.9b04048.

Videos for the experiments shown in Figure 5 (MP4, MP4)

Stimulus signal amplitude and frequency selection for impedance sensing, uniformity measurements over 169 connections with DI water, silicon oil, and o.c. condition, and multifrequency detection of seven different samples (PDF)

## AUTHOR INFORMATION

**Corresponding Author**

**Hanbin Ma** − CAS Key Laboratory of Bio-Medical Diagnostics, Suzhou Institute of Biomedical Engineering and Technology, Chinese Academy of Sciences, Suzhou 215163, P.R. China; ACXEL Tech Ltd., Cambridge CB4 0GA, U.K.; ◉ orcid.org/0000-0002-7629-2287; Email: mahb@sibet.ac.cn

**Authors**

**Chunjie Zhang** − CAS Key Laboratory of Bio-Medical Diagnostics, Suzhou Institute of Biomedical Engineering and Technology, Chinese Academy of Sciences, Suzhou 215163, P.R. China; School of Electronics and Information Engineering, Soochow University, 215006 Suzhou, P.R. China

**Yang Su** − CAS Key Laboratory of Bio-Medical Diagnostics, Suzhou Institute of Biomedical Engineering and Technology, Chinese Academy of Sciences, Suzhou 215163, P.R. China; ACXEL Tech Ltd., Cambridge CB4 0GA, U.K.

**Siyi Hu** − CAS Key Laboratory of Bio-Medical Diagnostics, Suzhou Institute of Biomedical Engineering and Technology,
Chinese Academy of Sciences, Suzhou 215163, P.R. China; ◉ orcid.org/0000-0002-0686-5182

**Kai Jin** − CAS Key Laboratory of Bio-Medical Diagnostics, Suzhou Institute of Biomedical Engineering and Technology, Chinese Academy of Sciences, Suzhou 215163, P.R. China

**Yuhan Jie** − ACXEL Tech Ltd., Cambridge CB4 0GA, U.K.

**Wenshi Li** − School of Electronics and Information Engineering, Soochow University, 215006 Suzhou, P.R. China

**Arokia Nathan** − ACXEL Tech Ltd., Cambridge CB4 0GA, U.K.

Complete contact information is available at:
https://pubs.acs.org/10.1021/acsomega.9b04048

**Author Contributions**

Y.S., W.L., A.N., and H.M. conceived the concept and experiments. C.Z. designed sensing circuit. C.Z. and Y.J. carried out the experiments and collected data. K.J. and S.H. performed impedance measurements and data analysis. H.M., C.Z., and Y.S. wrote the manuscript, and all authors reviewed and commented on the manuscript.

**Funding**

This work supported by the National Natural Science Foundation of China (grant no. 61701493), Policy Guidance project (International Science and Technology Cooperation) of Jiangsu Province of China (BZ2018040), Project Funded by China Postdoctoral Science Foundation (2019M651959), and Postdoctoral Research Funding Program of Jiangsu Province (2018K004B).

**Notes**

The authors declare the following competing financial interest(s): Two patents based on this research have been submitted. C.Z. is an M.S. student from the School of Electronics and Information Engineering, Soochow University, under supervision of W.L. C.Z. carried out all the work in H.M.'s lab in the CAS Key Laboratory of Bio-Medical Diagnostics, Suzhou Institute of Biomedical Engineering and Technology, Chinese Academy of Sciences, as a visiting scholar. Y.S. is a co-founder of ACXEL Tech Ltd. and a visiting associate professor in the Suzhou Institute of Biomedical Engineering and Technology, Chinese Academy of Sciences. A.N. is a co-founder of ACXEL Tech Ltd. and a chief technical officer of Cambridge Touch Technologies Ltd. H.M. is a professor in the Suzhou Institute of Biomedical Engineering and Technology, Chinese Academy of Sciences and a co-founder of ACXEL Tech Ltd.

## ACKNOWLEDGMENTS

We thank the technical support from ACXEL's engineering team.

## REFERENCES

(1) Moore, G. E. Cramming more components onto integrated circuits. Proc. IEEE 1998, 86, 82−85.

(2) Hefner, A. R., Jr.; Blackburn, D. L. An Analytical Model for the Steady-State and Transient Characteristics of the Power Insulated-Gate Bipolar-Transistor. Solid-State Electron. 1988, 31, 1513−1532.

(3) Nathan, A.; Ahnood, A.; Cole, M. T.; Lee, S.; Suzuki, Y.; Hiralal, P.; Bonaccorso, F.; Hasan, T.; Garcia-Gancedo, L.; Dyadyusha, A.; Haque, S.; Andrew, P.; Hofmann, S.; Moultrie, J.; Chu, D.; Flewitt, A. J.; Ferrari, A. C.; Kelly, M. J.; Robertson, J.; Amaratunga, G. A. J.; Milne, W. I. Flexible Electronics: The Next Ubiquitous Platform. Proc. IEEE 2012, 100, 1486−1517.

(4) Wheeler, A. R. Putting electrowetting to work. *Science* **2008**, *322*, 539–540.

(5) Abdelgawad, M.; Wheeler, A. R. The Digital Revolution: A New Paradigm for Microfluidics. *Adv. Mater.* **2009**, *21*, 920–925.

(6) Pollack, M. G.; Shenderov, A. D.; Fair, R. B. Electrowetting-based actuation of droplets for integrated microfluidics. *Lab Chip* **2002**, *2*, 96–101.

(7) Samiei, E.; Tabrizian, M.; Hoorfar, M. A review of digital microfluidics as portable platforms for lab-on a-chip applications. *Lab Chip* **2016**, *16*, 2376–2396.

(8) Swyer, I.; von der Ecken, S.; Wu, B.; Jenne, A.; Soong, R.; Vincent, F.; Schmidig, D.; Frei, T.; Busse, F.; Stronks, H. J.; Simpson, A. J.; Wheeler, A. R. Digital microfluidics and nuclear magnetic resonance spectroscopy for in situ diffusion measurements and reaction monitoring. *Lab Chip* **2019**, *19*, 641–653.

(9) Weigl, B.; Domingo, G.; LaBarre, P.; Gerlach, J. Towards non- and minimally instrumented, microfluidics-based diagnostic devices. *Lab Chip* **2008**, *8*, 1999–2014.

(10) Jiang, C.; Ma, H.; Hasko, D. G.; Nathan, A. Influence of polarization on contact angle saturation during electrowetting. *Appl. Phys. Lett.* **2016**, *109*, 211601.

(11) Mugele, F.; Baret, J. C. Electrowetting: From basics to applications. *J. Phys.: Condens. Matter* **2005**, *17*, R705–R774.

(12) Pollack, M. G.; Fair, R. B.; Shenderov, A. D. Electrowetting-based actuation of liquid droplets for microfluidic applications. *Appl. Phys. Lett.* **2000**, *77*, 1725–1726.

(13) Zeng, J.; Korsmeyer, T. Principles of droplet electro-hydrodynamics for lab-on-a-chip. *Lab Chip* **2004**, *4*, 265–277.

(14) Hadwen, B.; Broder, G. R.; Morganti, D.; Jacobs, A.; Brown, C.; Hector, J. R.; Kubota, Y.; Morgan, H. Programmable large area digital microfluidic array with integrated droplet sensing for bioassays. *Lab Chip* **2012**, *12*, 3305–3313.

(15) Kalsi, S.; Valiadi, M.; Tsaloglou, M. N.; Parry-Jones, L.; Jacobs, A.; Watson, R.; Turner, C.; Amos, R.; Hadwen, B.; Buse, J.; Brown, C.; Sutton, M.; Morgan, H. Rapid and sensitive detection of antibiotic resistance on a programmable digital microfluidic platform. *Lab Chip* **2015**, *15*, 3065–3075.

(16) Malic, L.; Brassard, D.; Veres, T.; Tabrizian, M. Integration and detection of biochemical assays in digital microfluidic LOC devices. *Lab Chip* **2010**, *10*, 418–431.

(17) Alistar, M.; Gaudenz, U. OpenDrop: An Integrated Do-It-Yourself Platform for Personal Use of Biochips. *Bioengineering* **2017**, *4*, 45.

(18) Fobel, R.; Fobel, C.; Wheeler, A. R. Dropbot: an open-source digital microfluidic control system with precise control of electrostatic driving force and instantaneous drop velocity measurement. *Appl. Phys. Lett.* **2013**, *102*, 193513.

(19) Ng, A. H. C.; Fobel, R.; Fobel, C.; Lamanna, J.; Rackus, D. G.; Summers, A.; et al. A digital microfluidic system for serological immunoassays in remote settings. *Sci. Transl. Med.* **2018**, *10*, No. eaar6076.

(20) Li, J.; Ha, N. S.; Liu, T.; van Dam, R. M.; Kim, C.-J. Ionic-surfactant-mediated electro-dewetting for digital microfluidics. *Nature* **2019**, *572*, 507–510.

(21) Shah, G. J.; Ohta, A. T.; Chiou, E. P.-Y.; Wu, M. C.; Kim, C. J. EWOD-driven droplet microfluidic device integrated with optoelectronic tweezers as an automated platform for cellular isolation and analysis. *Lab Chip* **2009**, *9*, 1732–1739.

(22) Barbulovic-Nad, I.; Au, S. H.; Wheeler, A. R. A microfluidic platform for complete mammalian cell culture. *Lab Chip* **2010**, *10*, 1536–1542.

(23) Barbulovic-Nad, I.; Yang, H.; Park, P. S.; Wheeler, A. R. Digital microfluidics for cell-based assays. *Lab Chip* **2008**, *8*, 519–526.

(24) Critchfield, F. E.; Gibson, J. A., Jr.; Hall, J. L. Dielectric Constant and Refractive Index from 20 to 35° and Density at 25° for the System Tetrahydrofuran Water. *J. Am. Chem. Soc* **1953**, *75*, 6044–6045.

(25) Malmberg, C. G.; Maryott, A. A. Dielectric Constant of Water from 0° to 100° C. *J. Res. Natl. Bur. Stand.* **1956**, *56*, 1–8.